\documentclass[a4paper,dvips2t,superscriptaddress,showkeys]{revtex4} 
\usepackage{epsfig,graphics,graphicx,amsmath,amssymb}
\usepackage{graphics}
\usepackage{amssymb}
\usepackage{color}
\usepackage{amsmath}
\usepackage{amsthm}
\usepackage{amsopn}
\def\uno{\mbox{1 \kern-.59em {\rm l}}}

\def\be{\begin{equation}}
\def\ee{\end{equation}}
\def\bea{\begin{eqnarray}}
\def\eea{\end{eqnarray}}

\begin{document}
\title{\bf\bf Mixed states for neutral current neutrino oscillation}

\author{ M. M. Ettefaghi }
\email{mettefaghi@qom.ac.ir}
\affiliation{Department of Physics, University of Qom, Ghadir Blvd., Qom 371614-6611, I.R. Iran} 
\author{ Z. Askari pour Ravari}%\email{amojavezi98@gmail.com}
\affiliation{Department of Physics, Islamic Azad University, North Tehran Branch Tehran, I.R. Iran}

\begin{abstract}
The theory of neutrino oscillation predicts that if both neutrino and antineutrino coming from $Z_0$ decay are detected, one can observe an oscillation pattern  between the corresponding detectors. This prediction is based on two properties; the neutrino-antineutrino pairs are produced coherently and they are detected with definite flavor in detectors. In this paper, we reanalyze this problem with considering some massive neutrinos which are mixed with the light neutrinos but they either participate incoherently or are decoupled in the production and detection processes.  In fact, neutrinos whose masses are larger than the upper bound on the mass uncertainty to be compatible with the coherence conditions (we will see it is about 1 keV) must be treated incoherently. Very heavy neutrinos whose masses are much larger than the neutrino energy in the neutrino production process are decoupled.
 Under these conditions, the created neutrino-antineutrino state as well as the states of detected neutrino and antineutrino is mixed. We see that the oscillation pattern cannot be observed for incoherent neutrinos and the standard oscillation pattern is recovered if the light neutrino masses are ignored in the production and detection processes. Moreover, since the $Z_0$ decay process is performed blindly with respect to flavors, the oscillating contributions in the event rates are independent of the $Z_0$ decay width.
\end{abstract} 

\maketitle

\section{Introduction}
Neutrino oscillation is one of the most interesting phenomena in quantum mechanics which has been experimentally established \cite{1}. The quantum approaches to neutrino oscillation are
based on the existence of nonzero and non-degenerate neutrino masses. However, the differences are so smaller than the energy uncertainty in the creation and detection processes that the neutrino mass eigenstate cannot be distinguished. This point is used in the quantum mechanics approach; the states of the created and detected neutrinos (well known as flavor eigenstates) are written as a coherent superposition of mass eigenstates \cite{2}. Of course, the states of neutrinos participating in weak interactions are not exactly identical to the flavor eigenstates \cite{3,2007}. But if the mass of neutrinos can be ignored, their weak interaction states can be considered as a flavor eigenstate.	Moreover, flavor neutrinos produced or detected in processes which involve
more than one neutrino cannot be separately described
by pure states, but require a density matrix description \cite{4}. However, they can be approximated with a density matrix of a pure state only when the differences of the neutrino masses are neglected in the interaction process. In most studies, light active neutrinos (standard model neutrino) have been considered. Therefore, their states are pure and the oscillation probabilities can be obtained in the framework of either quantum mechanics or quantum field theory \cite{44}. Meanwhile, the effects of mixing of the three standard light neutrinos with heavy neutrinos, which are either 
decoupled because their masses are much larger
than the maximum neutrino energy in the production and detection
processes or produced and detected
incoherently because their mass differences are larger than the related energy uncertainties, have been investigated in Ref. \cite{4}. In fact, the standard neutrino oscillation probability is recovered provided that the masses of light neutrino are ignored in the production and detection processes.

The neutrino and anti-neutrino state coming from a real or virtual $Z_0$ decay is a coherent superposition of either the flavor eigenstates or mass eigenstates. In fact, every flavor eigenstate as well as every mass eigenstate is created with the same probability. Therefore, one can write, in general, the state of the created neutrino-antineutrino as follows:
\be
|\nu_Z\rangle=\dfrac{1}{\sqrt{N_l}}\sum _{i=1}^{N_l}\left|  \nu _i\right\rangle \left|  \bar\nu _i\right\rangle=\frac{1}{\sqrt{N_l}}\sum_{\alpha=e,\mu,\tau,...}^{N_l}|\nu_\alpha\rangle|\bar{\nu}_\alpha\rangle,\label{eq3}
\ee
``..." denotes all other flavor states until the $N_l$'th one which all of them are created coherently during the $Z_0$ decay process. The second equality is satisfied provided that the mixing matrix is unitary. If we considered any other neutrinos either being created incoherently or being decoupled of electroweak interactions, the mixing matrix including only light coherent neutrinos would not be unitary. Furthermore, in the usual condition that only either neutrino or antineutrino in $|\nu_Z\rangle$ can be detected, the other one must be traced out. Therefore, the related density matrix is completely classical and it is impossible to observe usual neutrino oscillation in this condition. However, if both neutrino and anti-neutrino are detected in a coherent manner, a oscillation pattern can be observed between detectors \cite{5}. This problem has been restudied by considering the localization properties in Refs. \cite{6,7}. Two proper conditions play fundamental roles in obtaining the oscillation pattern; $|\nu_Z\rangle$ is a coherent mixture of neutrino-antineutrino pairs $\nu_i{\bar \nu}_i$ and they are detected with definite flavor. %Considering the neutrino-antineutrinos whose masses are larger than the coherence upper bound on neutrino mass uncertainty, which is about 1 keV as we will see in the next section, but smaller than their energies, we must add incoherently their corresponding states to the density matrix of the state given in Eq. (\ref{eq3}). 
Let us consider neutrinos-antineutrinos whose masses are smaller than their energy but larger than the coherence upper bound on neutrino mass uncertainty, which is about 1 keV as we will see in the next section.  In this case, their state must be added incoherently to the density matrix given in Eq. (\ref{eq3}).
Moreover, we assume that there exist heavy neutrino-antineutrinos whose masses are much larger than the neutrino-antineutrino energies coming from $Z_0$ decay. This
situation occurs, for example, in see-saw models \cite{seasaw1,seasaw2,seasaw3,seasaw4}.  These neutrinos and antineutrinos are decoupled but they might affect the oscillation pattern by mixing with active neutrinos. So, under these conditions, we face a new situation compared to the theoretical framework considered in Refs. \cite{5,6,7} and it is the scope of this paper.

In the next section, we give an appropriate quantum state describing both coherently and incoherently created neutrino-antineutrino due to the $Z_0$ decay process and take into account the corresponding time evolution. In the section \ref{333}, we consider two mechanisms for neutrino and antineutrino detection via charged current interactions; both of them are detected by scattering off nucleons and  neutrino is scattered off electron and antineutrino is done like before. Accordingly, an appropriate state is written for every case and we discuss on the corresponding oscillation probability. Finally, in the section \ref{444}, we summarize our results.

\section{Neutrino-antineutrino states due to $Z_0$ decay}
According to Eq. (\ref{eq3}), a pair of neutrino and anti-neutrino which are entangled and blind with respect to the flavor might be produced during a $Z_0$ decay process. In fact, we disregard the neutrino mass differences and use the unitarity of mixing matrix in Eq. (\ref{eq3}). Let us consider that the masses of neutrinos are so large that they cannot be ignored. Therefore, the neutrino state must be given by:
\be
|\nu_Z\rangle=\frac{1}{\sqrt{R_l^P}}\sum_{k\leq N_l}M_{kk}^P|\nu_k\rangle|\bar\nu_k\rangle,
\ee
where $R_l^P=\sum_{i\leq N_l}| M_{ii}^P|^2$ and $M_{ii}^P$ denote the amplitude of $Z_0$ decay into a neutrino-antineutrino pair with mass $m_i$. It is clear that if we ignore the mass difference of neutrinos, we reach Eq. (\ref{eq3}) that is the standard expression.

Now, if we consider in addition to $N_l$ neutrinos being produced coherently, there exist $N_h$ heavy neutrinos which are produced incoherently, the initial state is mixed and must be described by the following density matrix:
\be
\rho=\frac{1}{R_l}\sum_{k,k'\leq N_l}M_{kk}^P{M^P}^*_{k'k'}|\nu_k,\bar\nu_k\rangle\langle\nu_{k'},\bar\nu_{k'}|+\frac{1}{R_h^P}\sum_{k=N_l+1}^{N_l+N_h}|M_{kk}^P|^2|\nu_k,\bar\nu_k\rangle\langle\nu_k,\bar\nu_k|,\label{density}
\ee
in which
\be
R_h^P=\sum_{i=N_l+1}^{N_l+N_h}| M_{ii}^P|^2.
\ee
The first sentence of Eq. (\ref{density}) describes the state of neutrino-antineutrino produced coherently (it contains off-diagonal elements). The second sentence is related to heavy neutrino-antineutrinos whose mass differences are larger than the quantum-mechanical energy uncertainty and being produced incoherently. Indeed, from the relativistic energy momentum dispersion relation, the mass uncertainty of neutrino (antineutrino) can be estimated by $\sigma_{m^2}\simeq 2\sqrt{2}E\sigma_E$, where $E$ and $\sigma_E$ are energy and the energy uncertainty, respectively \cite{charge}. Given that the $Z_0$ interaction with environment is neglected, $\sigma_E$ is given by the $Z_0$ decay width. For instance, in the $Z_0$ rest frame we have $\sigma_{m^2}\sim(7\text{GeV})^2$. Therefore neutrino (antineutrino) mass eigenstates $\nu_i$ ($\bar \nu_i$) and $\nu_j$ ($\bar \nu_j$) are created incoherently provided that  $|m_i^2-m_j^2|>(7\text{GeV})^2$. However, to observe the oscillation phenomenon, the neutrino-antineutrino state must preserve its coherence until the detection processes. To explain the loss of coherence, we need to consider the localization properties of neutrinos and antineutrinos. Accordingly, they must be described by localized wave packets of width $\sigma_x$, which propagate with group velocities $v_g$ given by $v_g=\frac{\partial E}{\partial p}=\frac{p}{E}$. The coherence loss takes place during the time $t_{\text{coh}}$ when the overlaps of the wave packets of various mass eigenstate are diminished i.e.
	 \begin{equation}
	 	t_{\text{coh}}\simeq\frac{\sigma_x}{\Delta v_g},
	 \end{equation}
where $\Delta v_g$ is the group velocity difference of two mass eigenstates with masses $m_i$ and $m_j$:
\be
\Delta v_g=|\frac{p_i}{E_i}-\frac{p_j}{E_j}|\simeq 2\frac{|m_i^2-m_j^2|}{m_{Z_0}^2}.
\ee
Here, in the last step, we consider the $Z_0$ rest frame and use $\frac{m_{i(j)}^2}{m_{Z_0}^2}\ll1$ which is reasonable according to the above discussion. Therefore, the coherence length in the $Z_0$ rest frame is defined by
\be
x_\text{coh}\simeq v_g\frac{\sigma_x}{\Delta v_g}\simeq \frac{m_{Z_0}^2}{2\Gamma_{Z_0\rightarrow\nu\bar\nu}|m_i^2-m_j^2|},
\ee
in which we use $\sigma_x\simeq\sigma_p^{-1}\simeq (\frac{E}{p}\sigma_E)^{-1}\simeq (\frac{E}{p}\Gamma_{Z_0\rightarrow\nu\bar\nu})^{-1}$. Now, let us suppose the distance between two detectors to be about 1m. Therefore, the coherence of neutrino-antineutrino state is preserved up to this distance provided that $|m_i^2-m_j^2|<(1\text{keV})^2$. But in real conditions, the detector distance is much larger than this value, so the above bound is much more restricting. However, if we consider the $Z_0$ boson in flight, this restricting bound is relaxed a bit. In this case, the coherence length scales as $x_\text{coh}\rightarrow\gamma^3x_\text{coh}$, where $\gamma$ is the Lorentz factor \cite{charge}. For instance, if the $Z_0$ energy is about 1TeV, $x_\text{coh}$ becomes about 1km.

If we ignore the mass differences for the light neutrinos, the density matrix in Eq. (\ref{density}) is simplified as follows:
\be
\rho=\frac{1}{N_l}\sum_{k,k'\leq N_l}|\nu_k,\bar\nu_k\rangle\langle\nu_{k'},\bar\nu_{k'}|+\frac{1}{R_h^P}\sum_{k=N_l+1}^{N_l+N_h}|M_{kk}^P|^2|\nu_k,\bar\nu_k\rangle\langle\nu_k,\bar\nu_k|.\label{sdensity}
\ee
After propagation in plane wave approximation, the density matrix given in Eq. (\ref{density}) is transformed as follows:
\be\label{e6}
\rho(t,L;\bar t,\bar L)=\frac{1}{N_l}\sum_{k,k'\leq N_l}e^{-i(E_k-E_{k'})(t+\bar t)+i(p_k-p_{k'})(L+\bar L)}|\nu_k,\bar\nu_k\rangle\langle\nu_{k'},\bar\nu_{k'}|+\frac{1}{R_h^P}\sum_{k=N_l+1}^{N_l+N_h}|M_{kk}^P|^2|\nu_k,\bar\nu_k\rangle\langle\nu_k,\bar\nu_k|,
\ee
where $t$ and $\bar t$ are the neutrino and anti-neutrino travel time from the source to the corresponding detectors in the distances $L$ and $\bar L$, respectively. Here, we chose the $Z_0$ rest frame and $E_k$ and $p_k$ denote the energy and momentum of the $k$'th mass eigenstate. 
With  a realistic assumption, one can suppose that the light neutrinos are extremely relativistic. Therefore, their mass eigenstate energy and momentum are approximated by the following relations \cite{2000qm}:
\begin{equation}
	E_k\approx E+\xi\frac{m_k^2}{2E},
\end{equation}
\begin{equation}
	p_k\approx E-(1-\xi)\frac{m_k^2}{2E},
\end{equation}
where $E$ is the neutrino energy in the limit of zero mass and $\xi$ is a dimensionless quantity that can be
estimated from energy-momentum conservation in the production processes. In the case of $Z_0$ decay in the rest frame, $E$ and $\xi$ are $m_{Z_0}/2$ and 0, respectively. Thus, without  
losing the generality of the problem, one can write Eq. (\ref{e6}) as follows:
\be\label{e77}
\rho(L,\bar L)=\frac{1}{N_l}\sum_{k,k'\leq N_l}e^{-i\frac{\Delta m^2_{kk'}}{2E}(L+\bar L)}|\nu_k,\bar\nu_k\rangle\langle\nu_{k'},\bar\nu_{k'}|+\frac{1}{R_h^P}\sum_{k=N_l+1}^{N_l+N_h}|M_{kk}^P|^2|\nu_k,\bar\nu_k\rangle\langle\nu_k,\bar\nu_k|.
\ee
Here, we see that only the state corresponding to the coherent production has evolved during the propagation.

\section{Detection processes}\label{333}
As was said, both neutrino and antineutrino coming from a $Z_0$ decay process must be detected in order to observe intuitively oscillation pattern between detectors. Detection processes can usually be performed by an interaction involving one or two neutrinos.
When the detection process is done through neutrino scattering off nucleus via the charged current interaction, one neutrino is involved in the detection processes
\be\label{e7}
\nu_\alpha+D_I\rightarrow D_F+l_\alpha ^-,
\ee
\be\label{e8}
\bar\nu_\beta+\bar D_I\rightarrow \bar D_F+l_\beta ^+.
\ee 
As an example of two neutrinos participating in the detection process, let us consider neutrinos are detected via the following charged current process:
\be
\nu_\alpha+e^-\rightarrow \nu_e+l_\alpha ^-,\label{l1}
\ee
Of course, for non-electron neutrinos, this process is not used in practice for neutrino oscillation
experiments, because the neutrino energy threshold
is high (about 10.92 GeV) and the cross section is about
one thousand times smaller than that of the corresponding charged current scattering on neutron. 
In the case of the neutral current neutrino oscillation, where both neutrino and anti-neutrino are to be detected, the probability of oscillation, in the context of density matrix theory is given by
\be\label{e11}
P_{\alpha\beta}(L,\bar L)=tr[\rho(L,\bar L)\rho^D_\alpha\otimes\bar \rho^{\bar D}_\beta],
\ee
where $\rho^D_\alpha$ and $\bar \rho^{\bar D}_\beta$ are the density matrices of a detected neutrino with flavor $\alpha$ and a detected antineutrino with flavor $\beta$, respectively. Similar to the production process, we assume that in addition to $N_l$ light neutrinos (antineutrinos) coherently involving in the detection processes, $N_h-N_l$ heavy neutrinos (antineutrinos) also participate incoherently in detection processes. Also, it may be possible a mixing between light
neutrinos and very heavy neutrinos which are decoupled
because their masses are much larger than the maximum
energy in the corresponding process.  

We consider two conceivable situations for combining two detection processes:   
\begin{itemize}
\item We assume that the both detection processes are done through interaction with the nucleus (see Eqs. (\ref{e7}) and (\ref{e8})). Hence, the density matrix of the detected neutrino and antineutrino will be as follows: 
\be\label{e12}
\rho^D_\alpha=\frac{|M_\alpha^0(E)|^2}{{R_{l\alpha}^D}}\sum_{j,j'\leq N_l}U_{\alpha j}^*U_{\alpha j'}
|\nu_j\rangle \langle \nu_{j'}|+\frac{1}{{R_{h\alpha}^D}}\sum_{j=N_l+1}^{N_l+N_h}|U_{\alpha j}|^2| M_j^{D}|^2|\nu_j\rangle \langle \nu_{j}|,
\ee
and
\be\label{e13}
\bar\rho^{\bar D}_\alpha=\frac{|\bar M_\alpha^0(E)|^2}{{\bar R_{l\alpha}^{\bar D}}}\sum_{j,j'\leq N_l}U_{\alpha j}U_{\alpha j'}^*
|\bar \nu_j\rangle \langle \bar \nu_{j'}|+\frac{1}{{\bar R_{h\alpha}^{\bar D}}}\sum_{j=N_l+1}^{N_l+N_h}|U_{\alpha j}|^2|\bar M_j^{\bar D}|^2|\bar \nu_j\rangle \langle \bar \nu_{j}|,
\ee
respectively. Here, we have ignored the mass difference for light neutrinos, so 
\be
{R_{l\alpha}^D}=|M_\alpha^0(E)|^2\sum_{j\leq N_l}|U_{\alpha j}|^2
\ee
\be
{\bar R_{l\alpha}^{\bar D}}=|\bar M_\alpha^0(E)|^2\sum_{j\leq N_l}|U_{\alpha j}|^2
\ee
and for incoherently involving heavy neutrinos and anti-neutrinos, we have
\be
{R_{h\alpha}^D}=\sum_{j=N_l+1}^{N_l+N_h}|U_{\alpha j}|^2| M_j^{D}|^2,
\ee
\be
{\bar R_{h\alpha}^{\bar D}}=\sum_{j=N_l+1}^{N_l+N_h}|U_{\alpha j}|^2|\bar M_j^{\bar D}|^2.
\ee
In Eqs. (\ref{e12}) and (\ref{e13}), we see that if the incoherent neutrinos are not considered the detected states are pure even though we do not ignore the neutrino masses. Inserting the density matrix operators given in Eqs. (\ref{e77}), (\ref{e12}) and (\ref{e13}) in Eq. (\ref{e11}), one can obtain the oscillation probability as follows:
\bea\label{166}
P_{\alpha\beta}(L,\bar L)&=&\frac{|M_\alpha^0(E)|^2|\bar M_\beta^0(E)|^2}{N_l{R_{l\alpha}^D}{\bar R_{l\beta}^{\bar D}}}\sum_{k,k'\leq N_l}U^*_{\alpha k}U_{\beta k}U_{\alpha k'}U^*_{\beta k'}e^{-i\frac{\Delta m^2_{kk'}}{2E}(L+\bar L)}\nonumber\\
&+&\frac{1}{R_h^P{R_{l\alpha}^D} {\bar R_{h\beta}^{\bar D}}}\sum_{k=N_l+1}^{N_l+N_h}|U_{\alpha k}|^2|U_{\beta k}|^2|M_{kk}^P|^2|M_k^D|^2|\bar M_k^{\bar D}|^2.
\eea
The first sentence in this equation does not show the standard oscillation probability between flavors $\alpha$ and $\beta$ because the new mixing matrix elements are $U_{\alpha k}/\sqrt{\sum_{j\leq N_l}|U_{\alpha j}|^2}$ which do not constitute an unitary matrix. The second term is related to neutrinos that do not represent oscillation behavior since their production and detection are performed incoherently. 
Meanwhile, the rate of oscillation  observation can be written as follows:
\be\label{18}
{\cal R}_{\alpha\beta}(L,\bar L,E)\propto \int d\text{PS}  \,\,(R^D_\alpha)(\bar R^{\bar D}_\beta)\,P_{\alpha\beta}(L,\bar L), 
\ee
where the integration over $d\text{PS}$ denotes schematically the integration over the phase space. $R^D_\alpha$ and $\bar R^{\bar D}_\beta$ are the probability of detection processes for neutrino and antineutrino, respectively. Thus, they are given by:
\be
R^D_\alpha=\sum_k |U_{\alpha k}|^2|M_k^D|^2,
\ee 
and
\be\label{24}
\bar R^{\bar D}_\beta=\sum_k |U_{\beta k}|^2|\bar M_k^{\bar D}|^2,
\ee   
Ignoring the mass differences for light neutrinos, one can write:
\be
R^D_\alpha=|M_\alpha^0(E)|^2\sum_{k\leq N_l}|U_{\alpha k}|^2+\sum_{k=N_l+1}^{N_l+N_h} |U_{\alpha k}|^2|M_k^D|^2={R_{l\alpha}^D}+{R_{h\alpha}^D},
\ee
and
\be
\bar R^{\bar D}_\beta=|\bar M_\beta^0(E)|^2\sum_{k\leq N_l}|U_{\beta k}|^2+\sum_{k=N_l+1}^{N_l+N_h}|U_{\beta k}|^2|\bar M_k^{\bar D}|^2={{\bar R}_{l\beta}^{\bar D}}+{{\bar R}_{h\beta}^{\bar D}},
\ee
where $M_\alpha^0(E)$ and $\bar M_\beta^0(E)$ are the amplitudes of the detection processes for massless neutrino and antineutrino.
Therefore, using the probability of transition given in Eq. (\ref{166}) and above issues, one can write the event rate as follows:
\bea\label{rate1}
{\cal R}_{\alpha\beta}(L,\bar L,E)&\propto&\sigma_\alpha^0(E)P_{\alpha\beta}^{\text{eff}}(L,\bar L)\bar\sigma^0_\beta(E)+\frac{\sum_{k,k'=N_l+1}^{N_l+N_h}|U_{\alpha k}|^2|U_{\beta k'}|^2\sigma^k_\alpha\bar\sigma^{k'}_\beta}{(\sum_{j\leq N_l}|U_{\alpha j}|^2)(\sum_{j\leq N_l}|U_{\beta j}|^2)}P_{\alpha\beta}^{\text{eff}}(L,\bar L)\nonumber\\
&+&
\sum_{k=N_l+1}^{N_l+N_h}\left(\frac{|U_{\alpha k}|^2|\sigma^{k}_\alpha\bar\sigma^0_\beta}{\sum_{j\leq N_l}|U_{\alpha j}|^2}
+\frac{|U_{\beta k}|^2\sigma^0_\alpha\bar\sigma^{k}_\beta}{{\sum_{j\leq N_l}|U_{\beta j}|^2}}\right)P_{\alpha\beta}^{\text{eff}}(L,\bar L)\dots
\eea
where $\sigma_\alpha^0(E)$ ($\bar \sigma_\alpha^0(E)$) and $\sigma_\beta^k(E)$ ($\bar\sigma_\beta^k(E)$) are the detection cross sections for neutrino (antineutrino) with masses zero and $m_k$, respectively. Here, ``..." denotes all no oscillating terms which are related to the incoherent neutrinos. $P_{\alpha\beta}^{\text{eff}}$ is similar to the usual oscillation formula (without considering heavy neutrinos) which is given by
\be
P_{\alpha\beta}^{\text{eff}}=\frac{1}{N_l}\sum_{k,k'\leq N_l}U^*_{\alpha k}U_{\beta k}U_{\alpha k'}U^*_{\beta k'}e^{-i\frac{\Delta m^2_{kk'}}{2E}(L+\bar L)}.
\ee
Eq. (\ref{rate1}) is reduced to the usual expected oscillation pattern if the incoherent neutrinos are not considered. Moreover, the scattering of incoherent neutrinos and antineutrinos in detectors contribute in the observation rate of oscillation pattern. %the produced due to $Z_0$ decay ones do not. 
Given that the transition probability given by Eq. (\ref{166}), only the no oscillating terms, which are not written explicitly in Eq. (\ref{rate1}), are dependent on the incoherent neutrino-antineutrinos producing through the $Z_0$ decay process. 

\item As another possibility, we assume that neutrinos and anti-neutrinos are detected by processes (\ref{l1}) and (\ref{e8}), respectively.
In the neutrino detection process, an electron neutrino is produced though the interaction of incoming neutrino with an electron. Given that no coherent superposition of outgoing neutrino mass eigenstate is detected, the cross section of the process (\ref{l1}) is
the incoherent sum of the cross sections with the different
massive neutrinos in the final state
\be
\sigma(\nu_\alpha+e^-\rightarrow \nu_e+l_\alpha ^-)=\sum_i\sigma(\nu_\alpha+e^-\rightarrow \nu_i+l_\alpha^-),\label{sl1}
\ee
Therefore, we consider the creation of a neutrino with mass $m_j$ as a finial state and construct the corresponding density matrix operator similar to Eq. (\ref{e12})
\be
\rho^D_j=\frac{|{M^D}_{0,j}|^2}{{{R_e}_l^D}_{\alpha j}}\sum_{k,k'\leq N_l}U_{\alpha k}^*U_{\alpha k'}
|\nu_k\rangle \langle \nu_{k'}|+\frac{1}{{{R_e}_h^{ D}}_{\alpha j}}\sum_{k=N_l+1}^{N_l+N_h}|U_{\alpha k}|^2| {M^{D}}_{kj}|^2|\nu_k\rangle \langle \nu_{k}|,
\ee
where 
\be
{{R_e}_l^D}_{\alpha j}=|                          {M^D}_{0,j}|^2\sum_{k\leq N_l}|U_{\alpha k}|^2,
\ee
and 
\be
{{R_e}_h^D}_{\alpha j}=\sum_{k= N_l+1}^{N_l+N_h}|U_{\alpha k}{M^D}_{k,j}|^2.
\ee
Now, we should sum over mass eigenstates of the outgoing neutrino with coefficients $U_{e,j}$ in order to write the detection density matrix operator appropriate for the process Eq. (\ref{l1}). So we have
\bea
\rho^D&=& \sum_{j=1}^{N_l+N_h}|U_{ej}|^2\rho^D_j\nonumber\\
&=&\sum_{j=1}^{N_l+N_h}\frac{|U_{ej}|^2}{\sum_{j\leq N_l}|U_{\alpha j}|^2}\sum_{k,k'\leq N_l}U_{\alpha k}^*U_{\alpha k'}|\nu_k\rangle \langle \nu_{k'}|
+\sum_{j=1}^{N_l+N_h}\frac{|U_{ej}|^2}{{{R_e}_h^{ D}}_{\alpha j}}\sum_{k=N_l+1}^{N_l+N_h}|U_{\alpha k}|^2| {M^{D}}_{kj}|^2|\nu_k\rangle \langle \nu_{k}|,
\eea
where the mass differences are ignored for light neutrinos. It should be noted here that even if we do not consider incoherent neutrinos, the neutrino state detected through the charged current leptonic process is mixed provided that the difference in the mass of coherent neutrinos is not ignored.  
Since we consider that anti-neutrinos are detected by the mechanism given by Eq. (\ref{e8}), the state of detected antineutrinos is given by the density matrix operator given in Eq. (\ref{e13}). Therefore, according to Eq. (\ref{e11}), the transition probability is obtained as follows:
\bea
P_{\alpha\beta}(L,\bar L)=&&
\frac{1}{\sum_{j\leq N_l}|U_{\beta j}|^2}\sum_{j=1}^{N_l+N_h}\frac{|U_{ej}|^2}{\sum_{i\leq N_l}|U_{\alpha i}|^2}\sum_{k,k'\leq N_l}U^*_{\alpha k}U_{\beta k}U_{\alpha k'}U^*_{\beta k'}e^{-i\frac{\Delta m^2_{kk'}}{2E}(L+\bar L)}\nonumber\\
&&+\frac{1}{R_h^P{\bar R_{h\beta}^{\bar D}}}\sum_{j=1}^{N_l+N_h}\frac{|U_{ej}|^2}{{{R_e}_h^{ D}}_{\alpha j}}\sum_{k=N_l+1}^{N_l+N_h}|U_{\alpha k}|^2|U_{\beta k}|^2|{M^{D}}_{kj}|^2|M_{kk}^P|^2|M_k^{\bar D}|^2.
\eea
The probability of process in neutrino detector is given by
\be
R^D_{\alpha e}=\sum_{k,k'}|U_{\alpha k}|^2|U_{e k'}|^2|M^D_{k,k'}|^2.
\ee
Therefore, according to Eq. (\ref{18}), the event transition rate can be written
\bea\label{rate2}
{\cal R}_{\alpha \beta}(L,\bar L,E)\propto&& \sum_{j=1}^{N_l+N_h}|U_{ej}|^2\Bigg(\sum_{k\leq N_l}|U_{e k}|^2 \sigma^{0,0}_\alpha+\sum_{k=N_l+1}^{N_l+N_h}|U_{e k}|^2 \sigma^{0,k}_\alpha+\frac{\sum_{i\leq N_l}|U_{e i}|^2}{\sum_{i\leq N_l}|U_{\alpha i}|^2}\sum_{k=N_l+1}^{N_l+Nh}|U_{\alpha k}|^2\sigma^{k,0}_\alpha\nonumber\\
&&\hspace{2.2cm}+\frac{1}{\sum_{i\leq N_l}|U_{\alpha i}|^2}\sum_{k=N_l+1}^{N_l+Nh}\sum_{k'=N_l+1}^{N_l+Nh}|U_{\alpha k}|^2|U_{e k'}|^2\sigma^{k,k'}_\alpha\Bigg)\nonumber\\
  &&\hspace{1.8cm}\times\left(\bar \sigma^0_\beta+\sum_{i=N_l+1}^{N_l+Nh}\frac{|U_{\beta i}|^2\bar \sigma^i_\beta}{\sum_{i'\leq N_l}|U_{\beta i'}|^2}\right)P_{\alpha\beta}^{\text{eff}}+\dots,
\eea
where ``..." denotes all no oscillating terms which are related to the incoherent neutrinos. This relation is very similar to Eq. (\ref{rate1}). The differences are related to the existence an outgoing neutrino in the neutrino detector. In fact, we have treated a mixed state as detected neutrino even though the incoherent neutrinos are not considered. The sum over the outgoing  neutrino mass eigenstates has appeared for this reason.

\end{itemize}

\section{summary and conclusion}\label{444}
Although the neutrinos created in neutral interactions do not have a specific flavor, due to the entanglement between the neutrino and antineutrino, the pattern of neutrino oscillation between two detectors is expected to occur if both of them are detected \cite{5}. This expectation is based on the assumption that neutrinos-antineutrinos are produced coherently in these processes and that they are detected in flavor states. 
In this paper, we challenge both of these assumptions by considering heavy neutrino-antineutrinos that are either incoherently created and detected or, despite mixing with active ones, their mass exceeds the available energy and they do not participate in the creation and detection processes. The existence of such neutrinos is predicted in some models such as the seesaw model \cite{seasaw1,seasaw2,seasaw3,seasaw4}. It is clear that the incoherent part of the created neutrino-antineutrino state does not evolve with time. Using the plane wave approach, we reanalyzed the issue of neutral current neutrino oscillation and obtained the most general expression for the event rate for two combination of detectors; both neutrino and antineutrino are detected by charge current nucleon scattering without neutrino in final states and neutrinos are detected by charged current leptonic process involving outgoing neutrino but antineutrinos are detected as the previous case (please see Eqs. (\ref{rate1}) and (\ref{rate2}), respectively). For both cases, the detected states are mixed because we consider the massive neutrinos and antineutrinos involving incoherently in the corresponding detection processes. However, in case of the later, since there is an outgoing neutrino which does not detected, the detected state is a mixed state even though the incoherent neutrinos do not considered. But this state transforms to a pure state provided that we ignore neutrino masses in the scattering process.  In general, the standard oscillation pattern is recovered provided that the neutrino masses are ignored in the production and detection processes.  
As a main result, this study shows again that the oscillation pattern is predicted between neutrinos and antineutrinos with definite flavor states. So the $Z_0$ decay width which is blind with respect to flavor does not appear in the oscillating terms in the event rates.

\end{document}